\documentclass[prl,twocolumn,amsfonts,amsmath,tightenlines,preprintnumbers,nofootinbib,superscriptaddress,letterpaper]{revtex4}
\usepackage{graphicx}
\usepackage{bm}

\newcommand{\gsim}{\raisebox{-0.7ex}{$\stackrel{\textstyle >}{\sim}$ }}
\newcommand{\Hbind}{ 16.6 \pm 2.1 \pm 4.5 \pm 1.0 \pm 0.6~{\rm MeV} }
\newcommand{\Hbindabs}{ 16.6 \pm 2.1 \pm 4.6~{\rm MeV} }
\newcommand{\Hprob}{ 0.98 }
\newcommand{\Hbonesix}{ 12.3(1.1)(4.0) }
\newcommand{\Hbtwozero}{ 4.5(1.1)(1.3) }
\newcommand{\Hbtwofour}{ 16.3(1.2)(1.4) }
\newcommand{\Hbthreetwo}{ 16.6(1.4)(3.1) }

\begin{document}

\preprint{UNH-10-04}
\preprint{ICCUB-10-201}
\preprint{JLAB-THY-10-1296}
\preprint{NT@UW-10-26}
\preprint{IUHET-554}
\preprint{UCB-NPAT-10-003}

\begin{figure}[!t]
\vskip -1.1cm
\leftline{
\includegraphics[width=3.0 cm]{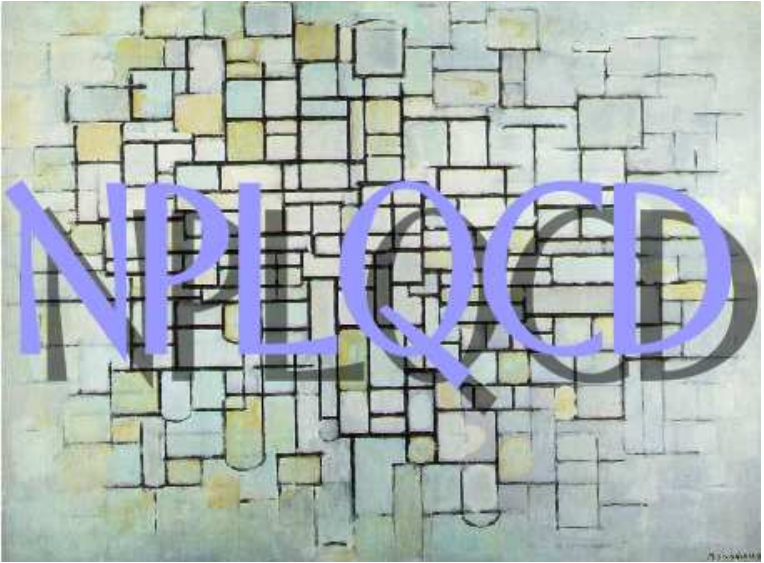}}
\vskip -0.5cm
\end{figure}

\title{Evidence for a Bound H-dibaryon from Lattice QCD}

\author{S.R.~Beane} 

\affiliation{Albert Einstein Zentrum f\"ur Fundamentale Physik,
Institut f\"ur theoretische Physik,
Sidlerstrasse 5,
CH-3012 Bern, Switzerland}
\affiliation{Department of Physics, University
  of New Hampshire, Durham, NH 03824-3568, USA}

\author{E.~Chang}
\affiliation{Dept. d'Estructura i Constituents de la Mat\`eria. 
Institut de Ci\`encies del Cosmos (ICC),
Universitat de Barcelona, Mart\'{\i} Franqu\`es 1, E08028-Spain}

\author{W.~Detmold} 
\affiliation{Department of Physics, College of William and Mary, Williamsburg,
  VA 23187-8795, USA}
\affiliation{Jefferson Laboratory, 12000 Jefferson Avenue, 
Newport News, VA 23606, USA}

\author{B.~Joo} 
\affiliation{Jefferson Laboratory, 12000 Jefferson Avenue, 
Newport News, VA 23606, USA}

\author{H.W.~Lin}
\affiliation{Department of Physics,
  University of Washington, Box 351560, Seattle, WA 98195, USA}

\author{T.C.~Luu}
\affiliation{N Division, Lawrence Livermore National Laboratory, Livermore, CA
  94551, USA}

\author{K.~Orginos}
\affiliation{Department of Physics, College of William and Mary, Williamsburg,
  VA 23187-8795, USA}
\affiliation{Jefferson Laboratory, 12000 Jefferson Avenue, 
Newport News, VA 23606, USA}

\author{A.~Parre\~no}
\affiliation{Dept. d'Estructura i Constituents de la Mat\`eria. 
Institut de Ci\`encies del Cosmos (ICC),
Universitat de Barcelona, Mart\'{\i} Franqu\`es 1, E08028-Spain}

\author{M.J.~Savage} \affiliation{Department of Physics,
  University of Washington, Box 351560, Seattle, WA 98195, USA}

\author{A.~Torok} \affiliation{Department of Physics, Indiana University,
  Bloomington, IN 47405, USA}
\author{A.~Walker-Loud}
\affiliation{Lawrence Berkeley National Laboratory, Berkeley, CA 94720, USA}

\collaboration{ NPLQCD Collaboration }

\date\today

\begin{abstract}
\noindent
We present evidence for the existence of a bound 
H-dibaryon, 
an $I=0$, $J=0$, $s=-2$ state with valence quark
structure $uuddss$,
at a pion mass of $m_\pi\sim 389~{\rm MeV}$.
Using the results of Lattice QCD calculations performed 
on four ensembles of anisotropic clover gauge-field configurations, 
with spatial extents
of $L\sim 2.0$, $2.5$, $3.0$ and $3.9~{\rm fm}$ at 
a spatial lattice spacing of $b_s\sim 0.123~{\rm fm}$,
we find an H-dibaryon bound by $B^{\rm H}_\infty =
\Hbindabs$ at a pion mass of $m_\pi\sim 389$ MeV.
\end{abstract}

\maketitle


\noindent 
It is now well established that quantum chromodynamics (QCD), the
theory describing the dynamics of quarks and gluons, and the
electroweak interactions, underlie all of nuclear physics, from the
hadronic mass spectrum to the synthesis of heavy elements
in stars.  To date, there have been few quantitative connections
between nuclear physics and QCD, but fortunately, Lattice QCD is
entering an era in which precise predictions for hadronic quantities
with quantifiable errors are being made.  This development is
particularly important for processes which are difficult to explore in
the laboratory, such as hyperon-hyperon and hyperon-nucleon
interactions for which knowledge is scarce, primarily due to the short
lifetimes of the hyperons, but which may impact the late-stages of 
supernovae evolution.  In this letter we report strong evidence
for a bound H-dibaryon, a six-quark hadron with valence structure
$uuddss$, from $n_f=2+1$ Lattice QCD calculations at light-quark
masses that give the pion a mass of $m_\pi\sim 389~{\rm MeV}$.

The prediction of a relatively deeply bound system with the quantum
numbers of $\Lambda\Lambda$ (called the H-dibaryon) by
Jaffe~\cite{Jaffe:1976yi} in the late 1970s, based upon a bag-model
calculation, started a vigorous search for such a system, both
experimentally and also with alternate theoretical tools.
Experimental constraints on, and phenomenological models of, the
H-dibaryon can be found in Refs.~\cite{Yamamoto:2000wf,Sakai:1999qm,Mulders:1982da}.
While experimental studies of doubly-strange hypernuclei restrict
the H-dibaryon to be unbound or to have a small binding energy,
the most recent constraints on the existence of the H-dibaryon come
from heavy-ion collisions at RHIC, from which it is concluded that the
H-dibaryon does not exist in the mass region 
$2.136 < M_{\rm H} < 2.231~{\rm
  GeV}$~\cite{Trattner:2006jn}, effectively eliminating the
possibility of a loosely-bound H-dibaryon at the physical light-quark
masses. 
Recent experiments at KEK suggest there is  
a resonance near threshold in the H-dibaryon
channel~\cite{Yoon:2007aq}.

The first study of baryon-baryon interactions with Lattice QCD was
performed more than a decade ago~\cite{Fukugita:1994na,Fukugita:1994ve}.  
This calculation was quenched and with $m_\pi\gsim 550~{\rm MeV}$.  
The NPLQCD
collaboration performed the first $n_f=2+1$ QCD calculations of
baryon-baryon interactions~\cite{Beane:2006mx,Beane:2006gf} at
low-energies but at unphysical pion masses. Quenched and dynamical
calculations were subsequently performed by the HALQCD
collaboration~\cite{Ishii:2006ec,Nemura:2008sp}. 
A number of quenched Lattice QCD
calculations~\cite{Mackenzie:1985vv,Iwasaki:1987db,Pochinsky:1998zi,Wetzorke:1999rt,Wetzorke:2002mx,Luo:2007zzb}
have searched for the H-dibaryon, but to date no definitive results
have been reported.  Earlier work concluded that the H-dibaryon does
not exist as a stable hadron in quenched QCD~\cite{Wetzorke:2002mx},
while more recent work~\cite{Luo:2007zzb,Beane:2009py} finds a hint of a bound
state.  
By inserting energy- and sink-dependent potentials into the Schr\"odinger
equation in the SU(3) limit, a hint of an H-dibaryon has been found in
Ref.~\cite{Inoue:2010hs}, however, this hint evaporates when
SU(3)-breaking is included~\cite{Inoue:2010xq}.

%
In this work, L\"uscher's
method~\cite{Hamber:1983vu,Luscher:1986pf,Luscher:1990ux,Beane:2003da}
is employed to extract two-particle scattering amplitudes below
inelastic thresholds from Lattice QCD calculations.  In the situation
where only a single scattering channel is kinematically allowed, the
deviation of the energy eigenvalues of the two-hadron system in the
lattice volume from the sum of the single-hadron masses is related to
the scattering phase shift, $\delta(q)$, 
as is made explicit in eq.~(\ref{eq:2}).  
The Euclidean time behavior of Lattice QCD correlation functions of the form 
$C_\chi(t)=\langle 0 |\chi(t)\chi^\dagger(0)|0\rangle$,
where $\chi$ represents an interpolating operator with the quantum numbers 
of the one-particle or two-particle systems under consideration, determines 
the ground state energies of the one-particle and two-particle 
systems, $E_1=m$ and $E_2$, respectively 
(we focus only on the ground state of the two-particle system in this work).  
The form of the interpolating operators, 
and the methodology used for extracting 
the energy shift are discussed in detail in Ref.~\cite{Beane:2010em}.  
For gauge-field configurations that have different lattice spacings in the
temporal and spatial directions (anisotropic lattices), 
the two-particle energy is 
given by $ E_2 = 2\sqrt{q^2/\xi^2 + m^2}$, 
where $\xi=b_s/b_t$ is the lattice anisotropy.  
By computing the mass of the particle and the ground-state energy 
of the two-particle system, the 
squared momentum, $q^2$ (in spatial lattice units (s.l.u)), 
which can be either positive or negative,
is determined by this relation.
For s-wave
scattering below inelastic thresholds, $q^2$ is related to the real
part of the inverse scattering amplitude through the eigenvalue 
equation~\cite{Luscher:1986pf}
(neglecting phase-shifts in $l\ge 4$ partial-waves)
\begin{equation}
  \label{eq:2}
  q\, \cot \delta(q) = 
  \frac{1}{\pi\ L} S\left(q^2 \left(\frac{L}{2\pi}\right)^2\right)\,,
\end{equation}
where the S-function is given by
\begin{equation}
  \label{eq:3}
  S(x)=\lim_{\Lambda\to\infty} \sum_{\bf j}^{|{\bf
      j}|<\Lambda}\frac{1}{|{\bf j}|^2 - x}  -4\pi\ \Lambda\,.
\end{equation}
This relation  provides a Lattice QCD determination of the value of 
the phase shift at the momentum $\sqrt{q^2}$.

Determining energy-levels with the same quantum numbers in multiple
volumes allows for bound states
to be distinguished from 
scattering states. 
A bound state corresponds to a pole in the S matrix, and 
in the case of a single scattering channel, is signaled by
$\cot\delta(q)\rightarrow +i$ in the large volume limit.
Writing $q = i\kappa$ for 
two-particle
states that are
negatively shifted in energy, $E_2<2 m$, 
in the lattice volume, the
volume dependence of the binding momentum in the large volume limit
follows directly from eq.~(\ref{eq:2}) and is of the
form~\cite{Beane:2003da}
\begin{eqnarray}
  \label{eq:4}
\kappa \ &\ =\ & 
\gamma \ +\ \frac{g_1}{L}\,\left(\ e^{-\gamma L}
\ +\ \sqrt{2}\  e^{-\sqrt{2}\gamma L}\ \right)\  +\ ... 
 \ ,
\label{eq:classicRES}
\end{eqnarray}
where $\gamma$ is the infinite-volume value of the binding momentum,
under the assumption that $\gamma\ll m_\pi$, and $g_1$ is treated as a
fit parameter.  With calculations in two or more  lattice volumes
that both have $q^2<0$ and $q\cot\delta(q)<0$ it is possible to
perform an extrapolation with eq.~(\ref{eq:classicRES}) to the
infinite-volume limit to determine the binding energy of the bound
state, $B_{\infty}=\gamma^2/m$.  The range of nuclear interactions is set by
the pion mass, and therefore the use of L\"uscher's method requires
that $m_\pi L\gg 1$ in order to strongly suppress the contributions
that depend upon the volume as $e^{-m_\pi L}$~\cite{Sato:2007ms}.

Our present results are from calculations on four ensembles of
$n_f=2+1$ anisotropic clover gauge-field configurations at a pion mass
of $m_\pi\sim 389$ MeV, a spatial lattice spacing of $b_s\sim
0.123(1)~{\rm fm}$, an anisotropy~\cite{Lin:2008pr,Edwards:2008ja} of
$\xi= 3.50(3)$, and with spatial-extents of $16,20,24,32$
lattice sites, corresponding to spatial dimensions of $L\sim 2.0$,
$2.5$, $3.0$ and $3.9~{\rm fm}$ respectively, and temporal extents of
$128$, $128$, $128$, and $256$ lattice sites, respectively.  The
precision of the calculations is sufficiently high that the
exponential volume dependence of the single baryon masses can be
cleanly quantified.
\begin{figure}[!th]
  \centering
  \includegraphics[width=0.48\columnwidth]{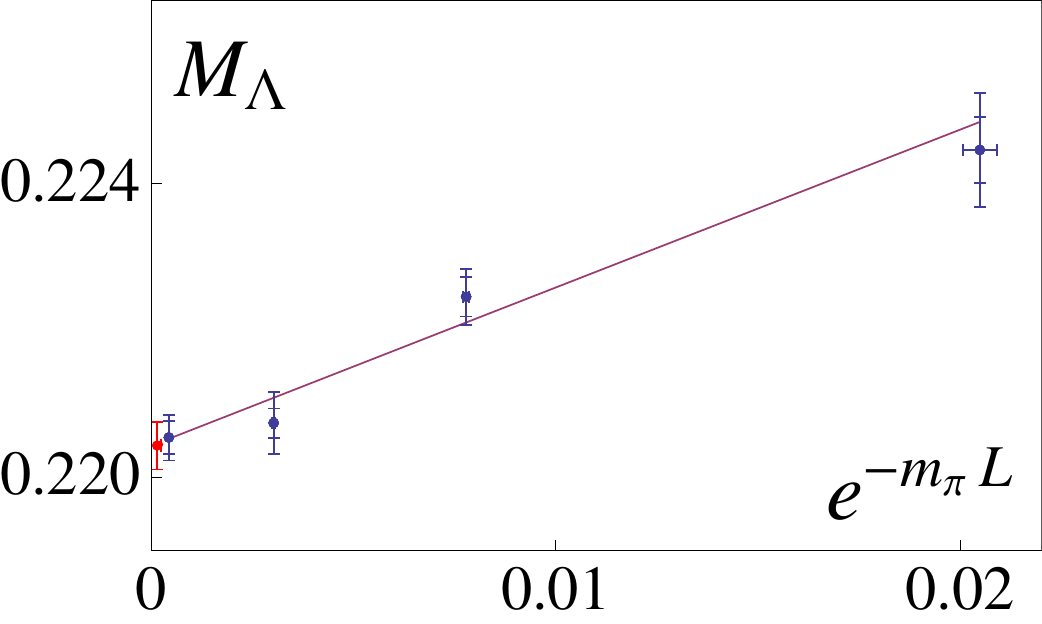}\ \ \ \ 
  \includegraphics[width=0.46\columnwidth]{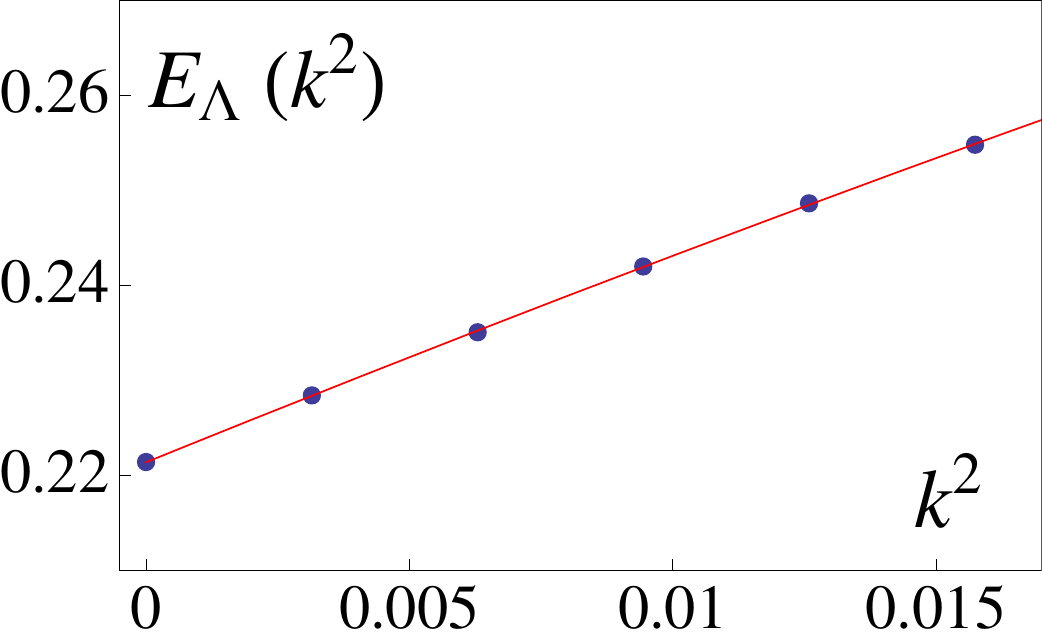}
  \caption{Left panel: the mass of the $\Lambda$ as a function of $e^{-m_\pi
      L}$ where $L$ is the spatial extent of the lattice.  
The left-most (red) point and uncertainty is the infinite-volume extrapolation
of the other (blue) points calculated in lattice volumes with 
spatial extents of, from left-to-right,  $L=32,24,20,$ and $16$.
The curve corresponds to the best straight-line fit.
    Right panel: the energy-momentum relation of the $\Lambda$ calculated 
on the $32^3\times 256$ ensemble. The
    points (and uncertainties) are the results of lattice 
    calculations and the (red) curve corresponds to the best fit (see text).
The units of the vertical axes in both plots 
are  ${\rm t.l.u.}$,
and of the horizontal axis of the right plot are  $({\rm t.l.u.})^2$
}
  \label{fig:Lamvol}
\end{figure}
The $\Lambda$ mass, unlike that of the $\pi$ and kaon, is found to
have statistically significant volume-dependence, as shown in the left
panel of fig.~\ref{fig:Lamvol}.  It is clear that the $\Lambda$ mass
on the $16^3\times 128$ ensemble ($m_\pi L~=~3.9$) is significantly
higher than its infinite-volume value and, more importantly, is
shifted by an amount that is comparable to the two-baryon energy
shifts.  The deviation found in calculations on the $20^3\times 128$
ensemble ($m_\pi L~=~4.8$) is much less than that of the $16^3\times
128$ ensemble, but we choose to use only calculations on the
$24^3\times 128$ ensemble ($m_\pi L~=~5.8$) and on the $32^3\times
256$ ensemble ($m_\pi L~=~7.7$) in the bound-state analysis.


L\"uscher's method assumes that the continuum single-hadron energy-momentum
relation is satisfied over the range of energies used in the eigenvalue
equation in eq.~(\ref{eq:2}).
In order to verify that 
this is the case,
single hadron correlation functions were formed with well-defined
lattice spatial momentum, ${\bf k}={2\pi\over L} {\bf n}$ for $|{\bf
  n}|^2 \le 5$.  As the low-lying states in the lattice volume have
energies that are small compared with the $\Lambda$ mass, it is
sufficient to determine the non-relativistic energy-momentum relation,
\begin{eqnarray}
E_\Lambda & = & M_0\ +\ { |{\bf k}|^2\over 2 M_1}
\ -\ { |{\bf k}|^4\over 8 M_2^3}\ +\ ...
\ \ \ .
\label{eq:NEep}
\end{eqnarray}
The $\Lambda$ energy as a function of momentum 
calculated on the $32^3\times 256$ ensemble
is shown in the right
panel of fig.~\ref{fig:Lamvol}, and yields $M_0, M_1, M_2$ of
$0.22135(10)(05)$, $0.2231(34)(13)$, $0.261(26)(04)~{\rm t.l.u}$,
respectively.  Clearly the special-relativity limit of $M_0=M_1=M_2$
is satisfied, but an uncertainty of $\sim 2\%$ is introduced into
$q^2$ from the uncertainties in the energy-momentum relation.
The use of relativistic or lattice dispersion relations leads to 
similar conclusions.

In the absence of interactions, the $\Lambda\Lambda$-$\Xi
N$-$\Sigma\Sigma$ system is expected to exhibit three low-lying states
as the mass-splittings between the single-particle states 
are (on the $32^3\times 256$ ensemble)
\begin{eqnarray}
2(M_\Sigma - M_\Lambda) & = & 0.01317(13)(19)~{\rm t.l.u}
\nonumber\\
M_\Xi + M_N - 2  M_\Lambda & = & 0.003397(61)(65)~{\rm t.l.u}
\ \ .
\label{eq:BBrestmasses}
\end{eqnarray}
However, if interactions generate a bound state, 
it is expected that the 
splitting between the ground-state and the two additional states will
be larger than estimates based upon the single baryon rest
masses.
\begin{figure}[!th]
  \centering
  \includegraphics[width=0.47\columnwidth]{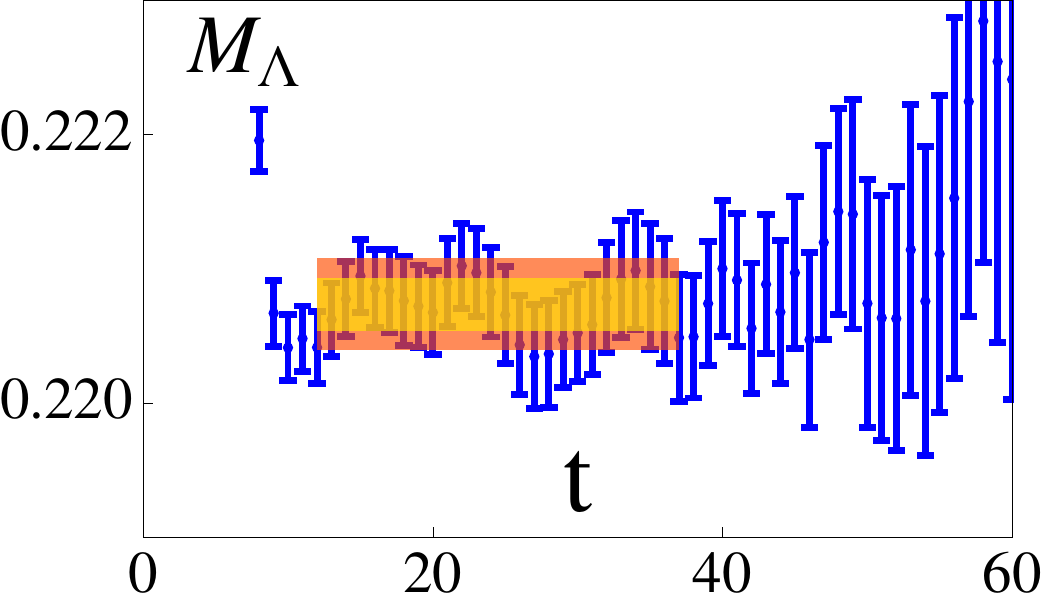}\ \ 
  \includegraphics[width=0.475\columnwidth]{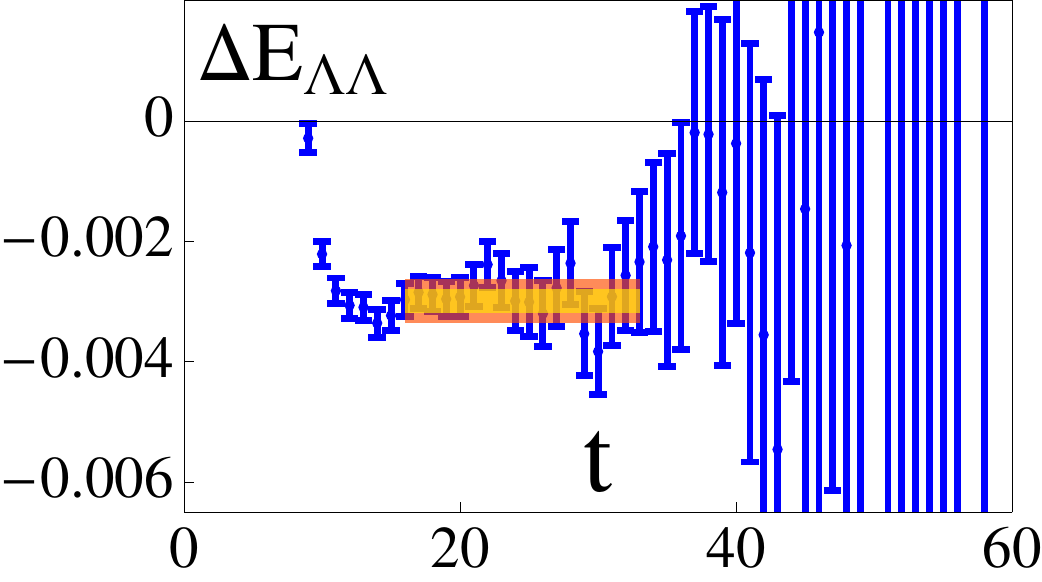}\ \ 
  \includegraphics[width=0.47\columnwidth]{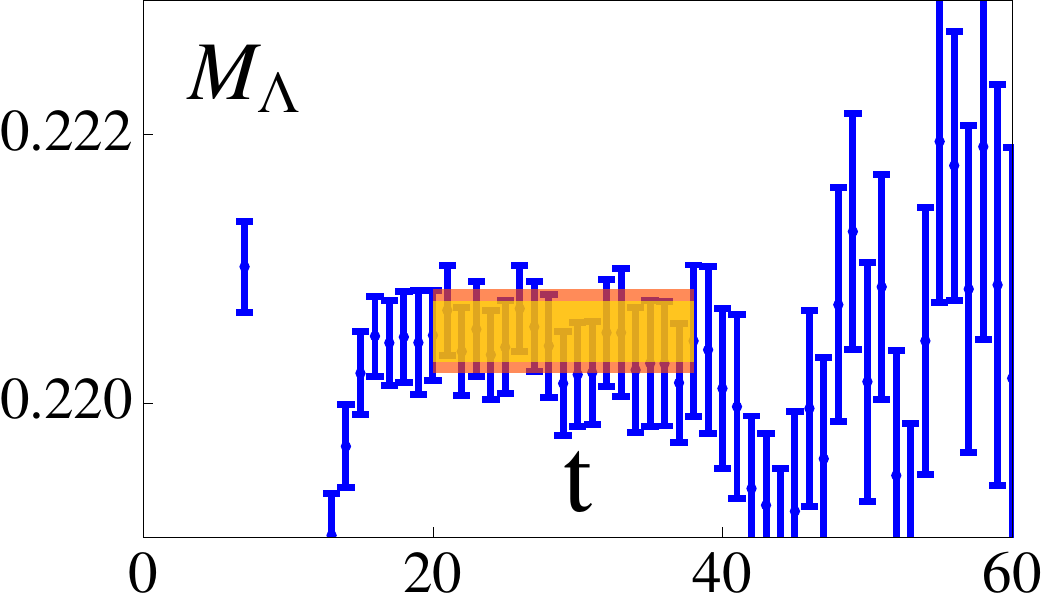}\ \  
  \includegraphics[width=0.475\columnwidth]{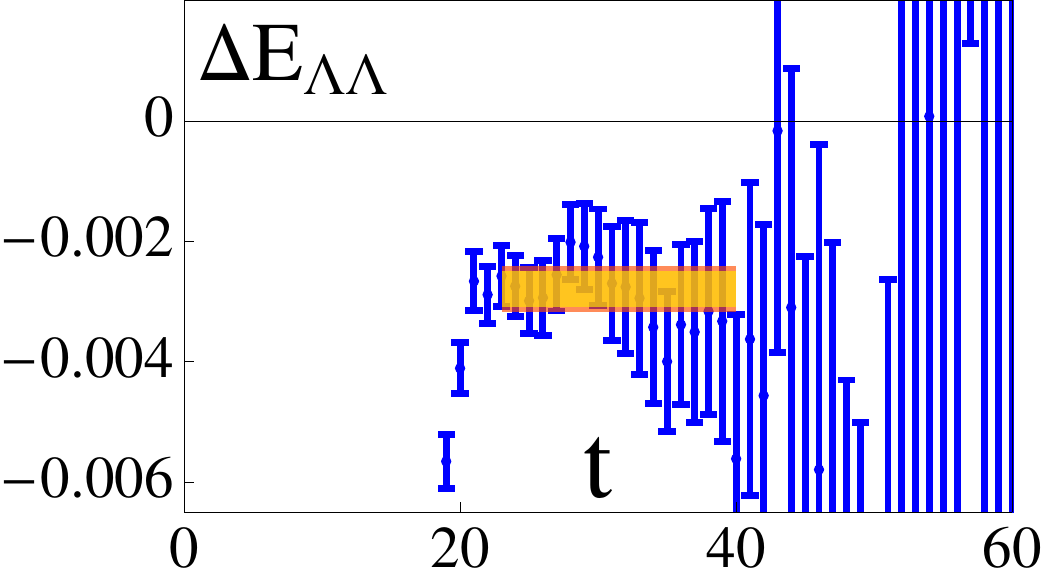}
  \caption{
The EMPs for the $\Lambda$ (left panels) and  the splitting between the 
$\Lambda\Lambda$ system and twice the $\Lambda$ mass (right panels) 
calculated on the 
$24^3\times 128$ (upper) and $32^3\times 256$ (lower) ensembles.  
The units of both axes are  t.l.u.
  }
\label{fig:EMPLam}
\end{figure}
The effective mass plot (EMP) for the $\Lambda$ calculated on the
$24^3\times 128$ and $32^3\times 256$ ensembles that have been
optimized for the ground-states using the matrix-Prony 
method~\cite{Beane:2009py} are shown
in the left panels of fig.~\ref{fig:EMPLam}, 
and clear plateaus are identified.  The calculated
EMP for the energy-splittings between the
$\Lambda\Lambda$-$\Xi N$-$\Sigma\Sigma$ coupled-channels (optimized using
the matrix-Prony method) and twice the energy of the $\Lambda$ (formed
from the ratio of correlation functions) on the $24^3\times 128$
and $32^3\times 256$ ensembles are shown in the right panels of
fig.~\ref{fig:EMPLam}.  
The finite-volume binding energies calculated
  in the $L=16$, $20$, $24$ and $32$ lattice volumes are $\Hbonesix$,
  $\Hbtwozero$, $\Hbtwofour$, and $\Hbthreetwo~{\rm MeV}$,
  respectively.
In each lattice
volume, the results are consistent with a single isolated
ground-state with an energy that is below the $\Lambda\Lambda$ threshold
(and considerably below the $\Xi N$ and $\Sigma\Sigma$ thresholds).
\begin{figure}[!th]
  \centering
  \includegraphics[width=0.95\columnwidth]{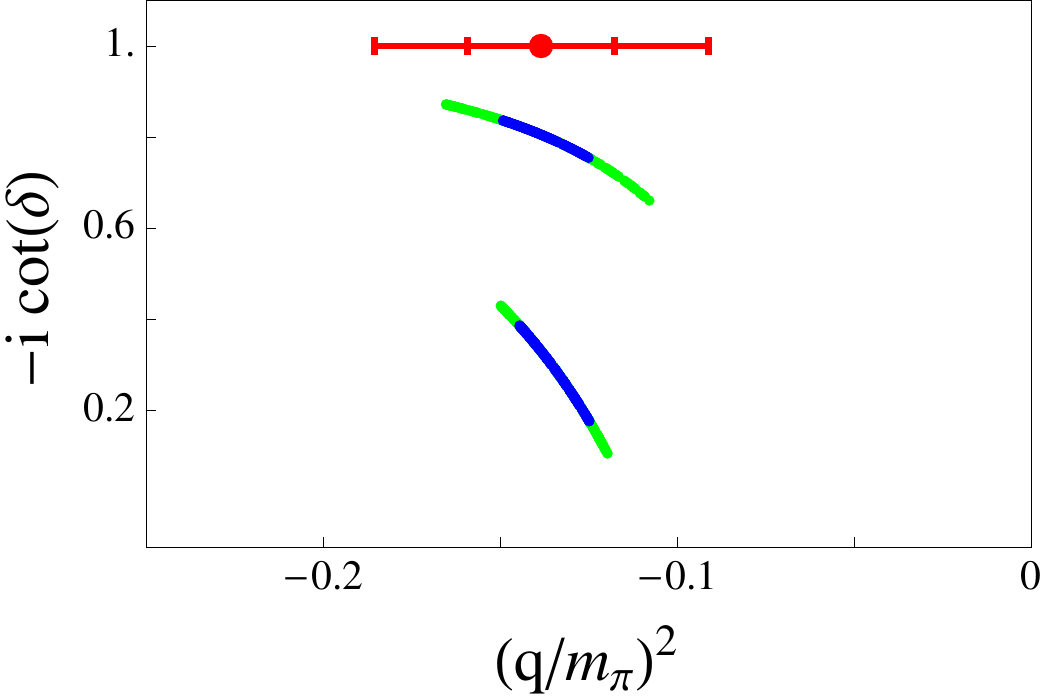}
  \caption{
The results of the Lattice QCD calculations of $-i\cot\delta$ 
versus $q^2/m_\pi^2$ obtained using eq.~(\protect\ref{eq:2}),
along with the infinite-volume extrapolation using eq.~(\protect\ref{eq:4}).  
The dark (blue) (light (green)) lines correspond to the statistical (systematic
and statistical uncertainties combined in quadrature)  $68\%$ confidence intervals
calculated on the $24^3\times 128$ ensemble (lower) and $32^3\times 256$
(upper) ensembles.
The (red) point and its uncertainty at 
$-i \cot\delta = +1$
corresponds to the 
infinite-volume extrapolation, the inner uncertainty  being statistical
and the outer  being the systematic
and statistical combined in quadrature.
  }
  \label{fig:extrapLamLam}
\end{figure}
The energy splittings and their uncertainties extracted from both
ensembles lead to negative values of $q \cot\delta$ indicating that
they both lie on the bound-state branch of the S-function (eq.~(\ref{eq:3})), 
and
thus leads us to identify the H-dibaryon.
The extracted values of
$-i\cot\delta$ from the $24^3\times 128$ and $32^3\times 256$
ensembles and their uncertainties
are shown in
fig.~\ref{fig:extrapLamLam},
along with the infinite-volume extrapolation implicit in eq.~(\ref{eq:2}), and 
made explicit in eq.~(\ref{eq:classicRES}).
The H-dibaryon binding energy at this
pion mass is found to be
\begin{eqnarray}
B^{\rm H}_\infty & = & \Hbind
\ \ ,
\label{eq:HBIND}
\end{eqnarray}
where the first uncertainty is statistical, the second systematic, the
third is an estimate of the uncertainty in the infinite-volume
extrapolation, and the fourth is the uncertainty from the
energy-momentum relation. 
Combining the various systematic uncertainties in quadrature gives
$B^{\rm H}_\infty = \Hbindabs$.
A Monte-Carlo propagation of the
uncertainties indicates that there is a probability greater than $\Hprob$ that the
H-dibaryon is bound at this pion mass.

In conclusion, we have presented strong evidence for the existence of
a bound H-dibaryon from Lattice QCD calculations at a pion mass of
$m_\pi\sim 389~{\rm MeV}$.  Our calculations were performed in four
lattice volumes, and a negatively shifted ground-state was found in
all four volumes. 
In order to 
avoid contamination from finite-volume modifications to
the $\Lambda$ mass and interactions, only the results obtained in the
larger two volumes were used in the infinite-volume extrapolation.
Within the uncertainties, the ground-state energies in the
largest two volumes are the same, indicating that both volumes are large
compared with the H-dibaryon size.  
This is consistent with the calculated binding energy.
Calculations were performed at only one lattice
spacing. However, given that lattice-spacing artifacts in these
calculations are expected to scale as ${\cal O}( b_s^2)$, we expect such
contributions to be small. 
Moreover, general arguments based on the
low-energy effective theory of the 
Symanzik action suggest that ${\cal O}( b_s^2)$ effects largely cancel
in forming the energy difference.  Consequently, we expect the observation
of the H-dibaryon to survive the continuum extrapolation.  
However, the quark-mass dependence of the H-dibaryon binding energy is presently
unknown, so a direct comparison of our result with experiment is not yet
possible.
As with all such lattice calculations, we cannot rule out the possibility of an
additional deeper bound state of the same quantum numbers in this channel 
that couples weakly to the interpolating operators.

The results of the Lattice QCD calculations presented in this letter provide
the first clear evidence for a
bound-state of two baryons directly from QCD.
This is further
strong motivation for pursuing Lattice QCD calculations in larger volumes, at
smaller lattice spacings, and over a range of light-quark masses
including those of nature, 
as the  present calculations demonstrate that the study of light (hyper-) 
nuclei directly from QCD is feasible.

\vskip 0.2in

We thank K. Roche for computing resources at ORNL NCCS
and R. Edwards
for help with QDP++ and Chroma~\cite{Edwards:2004sx}. 
We acknowledge computational support from the USQCD SciDAC project, NERSC (Office of
Science of the DOE, DE-AC02-05CH11231), the UW
HYAK facility, 
Centro Nacional de Supercomputaci\'on (Barcelona, Spain), LLNL,
and the NSF through Teragrid resources provided by
TACC and NICS under grant number TG-MCA06N025.  
SRB was supported in part by the NSF CAREER grant
PHY-0645570. 
The Albert Einstein Center for
Fundamental Physics is supported by the “Innovations- und
Kooperationsprojekt C-13” of the “Schweizerische
Universit\"atskonferenz SUK/CRUS”.  
The work of EC and AP is supported by the contract FIS2008-01661 
from MEC (Spain) and FEDER.
AP acknowledges support from the RTN Flavianet MRTN-CT-2006-035482 (EU).
H-WL and MJS were supported in part by
the DOE grant DE-FG03-97ER4014.  
WD and KO were
supported in part by DOE grants DE-AC05-06OR23177 (JSA) and 
DE-FG02-04ER41302. 
WD was also supported by DOE OJI grant
DE-SC0001784 and Jeffress Memorial Trust, grant J-968.  
KO was also supported in part by NSF grant CCF-0728915,
Jeffress Memorial Trust grant J-813 and
DOE OJI grant DE-FG02-07ER41527. 
AT was supported by 
NSF grant PHY-0555234 and DOE grant DE-FC02-06ER41443.
The work of TL was performed under the auspices of the U.S.~Department of
Energy by LLNL under Contract
DE-AC52-07NA27344 and the UNEDF SciDAC grant DE-FC02-07ER41457.
The work of AWL was supported in part by the Director, Office of Energy
Research, Office of High Energy and Nuclear Physics, Divisions of
Nuclear Physics, of the U.S. DOE under Contract No.
DE-AC02-05CH11231

%
%

\end{document}